# Interpretation of dielectric spectroscopy measurements of ferroelectric nematic liquid crystals

Vojko Matko[1], Ewa Gorecka[2], Damian Pociecha[2], Joanna Matraszek[2] and Nataša Vaupotič[3,4*]

[1] University of Maribor, Faculty of Electrical Engineering and Computer Science, Koroška 46, 2000 Maribor
[2] University of Warsaw, Faculty of Chemistry, Zwirki i Wigury 101, 02-089 Warsaw, Poland
[3] University of Maribor, Faculty of Natural Sciences and Mathematics, Koroška 160, 2000 Maribor, Slovenia
[4] Jozef Stefan Institute, Jamova 39, 1000 Ljubljana, Slovenia

*corresponding author: natasa.vaupotic@um.si

**Abstract**

The magnitude of the relative permittivity of the ferroelectric nematic phase ($N_F$) is under a lively scientific discussion since the phase was recently discovered. Dielectric spectroscopy measurements (DSM) give a huge value of relative permittivity, which depends on the cell thickness, but this is argued to result from a misinterpretation of the DSM results. We have conducted DSM using a set of cells differing in thickness of the $N_F$ layer, type of electrodes and presence/absence of nanoscale-thick surface layers. To model the DSM results, cells are presented by an equivalent electric circuit that includes a capacitor due to the $N_F$ layer with frequency dependent complex relative permittivity, capacitors due to surface layers, and a resistor describing limited conductivity of electrodes. DSM results for different cells with the same liquid crystal in the $N_F$ phase, are semi-quantitatively reproduced by the same set of physical parameters if a huge relative permittivity of the $N_F$, which is even orders of magnitude larger than the measured apparent values, is assumed. We show that the capacitance of surface layers should be considered also in cells with no polymer alignment layer on electrodes.

Ferroelectric nematic ($N_F$) liquid crystalline (LC) phase is a polar phase with the highest possible symmetry ($C_{\infty v}$). Its recent discovery [1,2] immediately attracted a wide research interest [3–8]. $N_F$ materials, although fluidic, are characterized by a strong order of dipoles resulting in a large spontaneous electric polarization of the order of $10^{-2}$ Cm$^{-2}$ [2,9,10], high values of the second order susceptibility and electrooptic coefficients [11–13], giant flexoelectric effect [14] and potentially phototunable permittivity [15]. These physical properties position ferroelectric nematics as promising candidates for diverse applications and are opening new roads to designing photonic structures [13]. Applications became even more promising with the experimental realization of chiral $N_F$ materials [16–19] and synthesis of materials having a $N_F$ phase at room temperature [9,20–22]. Another distinguishing feature of ferroelectric nematics is their remarkably high relative permittivity ($\sim 10^4$), within a frequency range up to the kHz regime [2,10,12,23–27]. This huge relative permittivity is attributed to the ease of thermally activated fluctuations of polarization direction. However, the relative permittivity in the $N_F$ phase appears to be strongly influenced by the thickness of cells and their surface treatment [23–27]. A similar effect is well-documented in ferroelectric smectic C [28] and orthogonal ferroelectric smectic A phases [29] and is explained by constrains of polar fluctuations by cell surfaces. However, because in the $N_F$ phase the effect is observed also in very thick cells, where surface interactions are negligible, this raised doubts about the interpretation of dielectric spectroscopy measurements (DSM) and posed a question on whether these materials are indeed characterized by large relative permittivity. Clark et al. suggested [30] that in liquid crystalline materials with very high spontaneous polarization, the large measured values of the relative permittivity are only apparent values, which result from the measured capacitance being misinterpreted as a capacitance of the whole cell, while the capacitance of very thin surface layers of surfactant is measured instead. Namely, the block reorientation of polarization in AC fields renders the $N_F$ layer effectively electrically conductive, which enables charging of very thin surface layers. The $N_F$ layer acts as a resistor with low ohmic resistance, its resistivity, $\rho_{LC}$, being dependent on the material rotational viscosity $\gamma$ and polarization $P$ as $\rho_{LC} = \gamma/P^2$. The relaxation time measured in DSM is that of a series connection of surface capacitors (very thin surface layers) and low-value resistor ($N_F$ layer) and the mode observed by DSM is called polarization-external capacitance Goldstone reorientation mode (PCG).

Proposition of the PCG mode set several research groups into further research of a proper interpretation of the DSM results of the ferroelectric nematics. Vaupotic et al. [23] pointed out that in the case of the relative permittivity being very large, the capacitance of the LC layer becomes comparable to the capacitance of surface layers, thus all capacitors are charged. In this case the interpretation of the DSM results becomes very complex (as noticed also by Yadav et al. [27]), so they focused on DSM in cells with bare gold electrodes (no surface layers of surfactant). They performed DSM for cells in a bias DC electric field and the behavior of the observed low frequency mode was explained by a continuous phenomenological model (CPM), which couples fluctuations of the nematic director and polarization. Erkoreka et al. [24] performed a systematic DSM as a function of cell thickness and discussed the results within both, the PCG and CPM, models. Clark et al. [30] interpreted results from Ref. [24] as an evidence of the PCG model, because it was observed that the low frequency impedance of the studied cells does not to depend on cell thickness. However, even though the dependence on the cell thickness is weak, it is there. By further investigations [25], Erkoreka et al. observed also high frequency relaxation processes predicted by CPM, thus confirming that relaxation processes within the $N_F$ phase are essentially detected by DSM.

During the refereeing process of the present letter, Adaka et al. [31] reported on a performed DSM in a set of cells, varying the thickness of the $N_F$ layer and thickness of the surface layer of surfactant. They found the apparent relative permittivity as well as the relaxation frequency being linearly dependent on the ratio between the thickness of the surface layer and the $N_F$ layer, as predicted by the PCG model. While the simplest (linear) dependence is an obvious first choice, a closer inspection of Fig. 2 in Ref. [31] shows that the experimental data, especially the dependence of the relaxation frequency, could also be fitted by a nonlinear dependence.

In this letter, we present DSM on a set of cells, differing in thickness, type of electrodes and presence/absence of a thin surface layer of surfactant. As mentioned above, the DSM results reported in Ref. [24] show that the low-frequency impedance of the studied cells exhibits dependence on the cell thickness. We also observed that cells with different thickness, but the same type of electrodes and thin surface layer have different low-frequency capacitance, contrary to the prediction of the PCG model. Moreover, analysis of our DSM results within the PCG model framework gives rather large resistance of the $N_F$ layer (leading to very large rotational viscosity), thus we decided to focus on the possibility of $N_F$ having a large relative permittivity, as, in fact, expected for a nonconductive ferroelectric material. This large value is due to the ease of polarization reorientation in the $N_F$ layer, while the layer is treated as an insulator. This approach is further supported by the observation of two low frequency dielectric modes in bias electric fields [23,29], suggesting that by DSM the frequency response of the $N_F$ is measured as well, not only charging and discharging of the thin surface layer.

Measurements were performed by using a standard material (analogue of RM734 [1], see **I-5** in ref. [18]) exhibiting a monotropic nematic and $N_F$ phases (m.p. 130.4°C, Iso 60.0°C N 52.7°C $N_F$). Material was filled in cells (planar capacitors) with gold or indium tin oxide (ITO) electrodes. Cells with gold electrodes were without polymer surface layers, while cells with ITO electrodes were either without or with surface layer of surfactant enforcing homeotropic anchoring. Information on the used cells is collected in **Table 1**. To make sure that systematic tendencies are observed, measurements were also performed for another ferroelectric nematic material (see Supplemental Material, SM).

**Table 1.** Used cells, thickness ($d$ [$\mu$m]), capacitance of empty cells ($C_0$ [pF]) and electrode surface area ($S$ [mm$^2$]). Low frequency (at 10 Hz) capacitances ($C$ [$\mu$F]) of cells and capacitances per unit surface area ($C/S$ [nFmm$^{-2}$]) for material in the $N_F$ phase at 49.0°C. Electrodes: gold with no surfactant layer (G) and ITO, with no surfactant layer (NA) and with surface layer of surfactant enforcing homeotropic anchoring (HT).

| cell | $d$ | $C_0$ | $S$ | $C$ | $C/S$ |
|------|-----|-------|-----|-----|-------|
| G5   | 5.0 | 44    | 25  | 0.32 | 13   |
| G10  | 10.0 | 22   | 25  | 0.25 | 10   |
| NA3  | 3.0 | 485   | 164 | 2.5 | 15    |
| NA5  | 4.9 | 95    | 53  | 0.52 | 9.8   |
| NA10 | 9.7 | 150   | 164 | 1.9 | 12    |
| HT10 | 10.0 | 77   | 87  | 0.29 | 3.3   |
| HT20 | 20.0 | 41.5 | 94  | 0.18 | 1.9   |

To obtain the frequency ($\nu$) dependence of the real and imaginary part of impedance of the cell filled with material in the $N_F$ phase, we measured voltage on the cell and phase shift between the voltage across the cell and the voltage applied by generator (Bode plot method, BPM) in a frequency ($\nu$) range from 10 Hz to 1 MHz. Details on the method are given in SM, where we also give a procedure for calculating the cell impedance ($Z$) and phase angle ($\alpha$) between the voltage across and current through the cell. For signal generation, Siglent SDG6022X 200MHz Function / Arbitrary Waveform Generator (16 bit) was used and measurements were recorded by a Siglent SDS6054A 4CH 500MHz 5 GSa/s Oscilloscope (16 bit). The generator output voltage was set to 100 mVpp for all the cells, much below the Frederick's transition voltage. Generator voltages from 10 mVpp to 200 mVpp were tested, as well, and BPM results differed only in the amount of noise at frequencies below 100 Hz (larger noise at lower amplitudes). Cells were heated by using PCB Heating platform HPB100 from iTECH and temperature was measured by PT100 temperature sensor and was stabilized with an accuracy of 0.1 K. The Siglent oscilloscope has 4 channels, one was used to measure the voltage output of function generator and the rest to measure simultaneously three cells, for which experimental conditions were thus identical.

The impedance $Z$ and phase angle $\alpha$ calculated from the BPM measurements in the $N_F$ phase are given in **Figure 1a,b**, raw data is given in SM (**Figure S2**). While $Z$ and $\alpha$ are obtained from the measured data without making any assumptions on the equivalent electric circuit of the cell, assumptions are needed to obtain the real and imaginary part of complex relative permittivity of the $N_F$ phase. To get some information on a proper equivalent electric circuit, we first interpret the measured $Z$ and $\alpha$ as if resulting from a capacitor with capacitance $C_S$ and resistor with resistance $R_S$ in a series connection (**Figure 1c,g**). These results can be analyzed within the scope of the PCG model, because the simplest equivalent circuit for the PCG model is a series connection of a capacitor due to thin surface layers ($C_S$) and a resistor due to the $N_F$ layer ($R_S$). We observe that the values of $C_S$ and $R_S$ are frequency dependent, moreover, $R_S$ values determined from experimental data are very high, up to $R_S \sim 10$ kΩ for the G10 cell. With typical values (by order of magnitudes) of polarization and rotational viscosities, $P \sim 5 \cdot 10^{-2}$ Cm$^{-2}$ and $\gamma \sim 1$ Pas, and the surface area ($S$) and thickness ($d$) of the G10 cell (see Table 1), we estimate that within the PCG model the resistance of the $N_F$ layer should be $\gamma d/(SP^2) \sim 160$ Ω. However, the experimental value for the G10 cell ($R_S \sim 10$ kΩ) could be modeled only by an extremely high value of rotational viscosity (of the order of $10^2$ Pas). In addition, the PCG model predicts that one should measure the same capacitance per unit surface of the capacitor for all cells with the same type of electrodes and surface alignment layer, because the capacitance should depend only on the thickness of surface layers. As seen in **Table 1**, the measured low frequency capacitances per unit surface area differ, but not in the ratio of cell thicknesses.

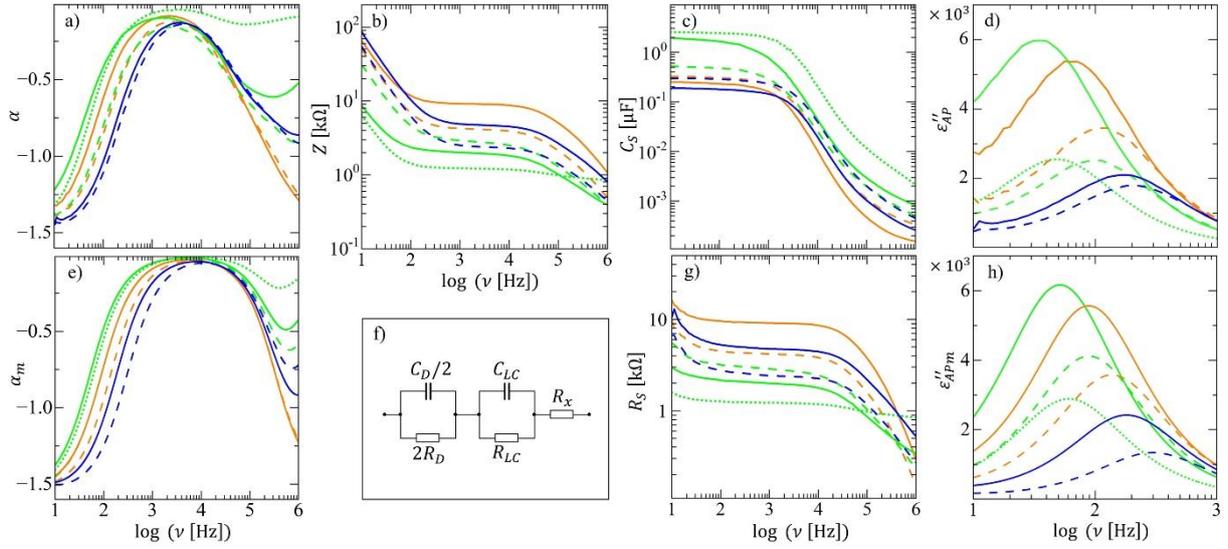

**Figure 1.** Frequency ($\nu$) dependence of a) phase angle ($\alpha$) between the current through and voltage across the cell and b) impedance ($Z$) calculated from the BPM measurements for the studied material in the N$_F$ phase at 49.0°C. Apparent c) capacitance ($C_S$) and g) resistance ($R_S$) of the cell obtained by simulating the measured $Z$ and $\alpha$ as resulting from a series connection of an ideal resistor and capacitor. d) Apparent imaginary part of complex permittivity ($\varepsilon''_{AP}$) calculated from the measured data by using eq. (3). f) Equivalent electric circuit of the cell filled with N$_F$; $C_D$ is capacitance of one surface layer and $R_D$ its resistance, $C_{LC}$ is capacitance and $R_{LC}$ resistance of the N$_F$ layer, $R_x$ is parasitic resistance. Simulated values of e) $\alpha_m$ and h) $\varepsilon''_{APm}$, calculated for the equivalent circuit given in f); material parameter values $\varepsilon_s = 2.5 \cdot 10^4$, $\tau = 4$ ms, $\varepsilon_\infty = 3$ and $\rho_{LC} = 10^7$ Ωm are the same for all cells; $R_D = 1$ MΩ. Values of $C_D$ and $R_x$ depend on the type of cells and are given in the main text. Green: NA10 (solid), NA5 (dashed), NA3 (dotted). Orange: G10 (solid), G5 (dashed). Blue: HT20 (solid), HT10 (dashed).

Based on the above observations we searched for an alternative model to explain results of DSM. Because the ratio of the low frequency capacitances does not equal the ratio of the cell thicknesses (see Table 1), we conclude that the influence of the thin surface layers cannot be neglected, not even in cells with bare electrodes, which is in line with the argument by Clark et al. [30] that a thin layer of liquid crystal molecules close to an electrode acts in the same way as a thin layer of surfactant. This stimulated us to revisit and modify the model equivalent circuit which considers both the capacitance of surface layers and capacitance of the N$_F$ layer (see Appendix in Ref. [23]). The proposed equivalent electric circuit is given in **Figure 1f**. The capacitance of one thin layer of surfactant (or thin layer of LC molecules) is $C_D$ and its resistance $R_D$; the net capacitance of two surface layers is $C_D/2$ and their net resistance $2R_D$. This resistance is important only at very low frequencies (lower than 10 Hz) and is in the model only for the sake of completeness. In general, $R_D$ can be assumed to be very large and as such can be omitted from the model equivalent circuit. We set $R_D$ to a constant value of 1 MΩ.

The capacitance of the N$_F$ layer ($C_{LC}$) is expressed as $C_{LC} = C_0\,\varepsilon(\omega)$. $C_0$ is capacitance of an empty cell and $\varepsilon(\omega)$, where $\omega = 2\pi\nu$, is a complex relative permittivity, which we approximate by the Debye relaxation model:

$$\varepsilon(\omega) = \varepsilon'(\omega) + i\varepsilon''(\omega) = \varepsilon_\infty + \frac{\varepsilon_s - \varepsilon_\infty}{1 + i\omega\tau}, \qquad (1)$$

where $\varepsilon'$ and $\varepsilon''$ are the real and imaginary part of the complex relative permittivity, respectively; $\varepsilon_\infty$ is a high-frequency and $\varepsilon_s$ low-frequency permittivity and $\tau$ is the Debye relaxation time. More general Havriliak – Negami equation [32] could be used instead, but to show the principle and have a lower number of fitting parameters we decided on the simplest relaxation model. The resistance of the liquid crystalline material is $R_{LC} = \rho_{LC}d/S$. The resistor with resistance $R_x$ takes account of parasitic resistances of electrodes and wiring. We point out that, except for the $R_x$ and $R_D$, the latter being irrelevant at $\nu > 10$ Hz, the equivalent circuit in **Figure 1f** is the same as the one used in the PCG model (see Fig. 1 in Ref. [30]). The difference is in the value of N$_F$ resistivity, $\rho_{LC}$, and the treatment of $C_{LC}$. In our model, $\rho_{LC}$ is large (characteristics of insulators) and it is mainly due to ionic impurities. In the PCG model, $\rho_{LC}$ is low and it is due to the block reorientation of polarization. In the here-proposed model the reorientation of polarization in the bulk N$_F$ layer is accounted for in the complex relative permittivity of N$_F$.

The complex impedance ($Z_c$) of the equivalent electric circuit (**Figure 1f**) is

$$Z_c = R_x + \left(\frac{1}{2R_D} + \frac{i\omega C_D}{2}\right)^{-1} + \left(\frac{1}{2R_{LC}} + i\omega C_{LC}\right)^{-1}, \qquad (2)$$

which can be expressed as a sum of the real ($Z_r$) and imaginary ($Z_i$) part of complex impedance as $Z_c = Z_r + iZ_i$, the expressions for $Z_r$ and $Z_i$ being functions of $\varepsilon'$ and $\varepsilon''$. When the circuit (cell) is connected to an AC voltage generator ($U(\omega) = U_{cell}e^{i\omega t}$, $U_{cell}$ being the amplitude of the voltage on the cell), the current $I(\omega) = U(\omega)Z_c^{-1} = I_0(\omega) + iI_{\pi/2}(\omega)$ through the circuit can be divided into the current in phase with voltage ($I_0 = U_{cell}Z_r|Z_c|^{-2}$) and current that is phase shifted by $\pi/2$ ($I_{\pi/2} = U_{cell}Z_i|Z_c|^{-2}$). The phase angle between the current and voltage is given by $\tan\alpha_m = Z_iZ_r^{-1}$, where the subscript $m$ next to $\alpha$ is used to differentiate the model phase angle from the one obtained from DSM. For $C_D \gg C_{LC}$, $R_D$ and $R_{LC}$ being large and $R_x$ small, the complex impedance reduces to $Z_c \approx i\omega C_{LC}$, thus only the LC capacitor is charged when the cell is connected to an AC voltage generator. In this case is it straightforward to see that the imaginary part of the complex permittivity is related to the current in phase with the voltage across the cell as $\varepsilon'' = I_0/(\omega C_0 U_{cell})$ and the current that is phase shifted by $\pi/2$ is related to the real part of complex permittivity as $\varepsilon' = I_{\pi/2}/(\omega C_0 U_{cell})$. However, when $C_D$ and $C_{LC}$ are comparable, $I_0/(\omega C_0 U_{cell})$ becomes a complex function of $C_0$, $\varepsilon'$, $\varepsilon''$, $R_D$, $R_{LC}$, $C_D$ and $R_x$, so we call this ratio an apparent value of the imaginary part of the complex relative permittivity ($\varepsilon''_{AP,m}$):

$$\varepsilon''_{AP,m} = \frac{I_0}{\omega C_0 U_{cell}}. \qquad (3)$$

In DSM, $\varepsilon''$ of the studied material is usually calculated by using the expression on the right-hand side of eq. (3), which means that only an apparent value of the imaginary part of the complex relative permittivity () is obtained. The same argument goes for the real part of complex permittivity.

Our goal was to find such values of model parameters that would give at least a semi-quantitative fit with the DSM results. The major objective was to find a set of material parameters ($\varepsilon_s$, $\tau$, $\varepsilon_\infty$) that are common for all the cells, regardless of their thickness and presence/type of thin layer of surfactant. In search for this set of parameters we noticed that physical quantities that are most sensitive to changes in parameter values are $\alpha(\omega)$ and $\varepsilon''_{AP}$.

The $\varepsilon''_{AP}$ calculated from the measured data by using eq. (3) is given in **Figure 1d**. The maximum value of $\varepsilon''_{AP}$ strongly depends on the cell thickness and type of the cell (NA, G or HT). Within the proposed model, this dependence is a consequence of a huge value of the low-frequency relative permittivity ($\varepsilon_s$) of N$_F$, due to which the condition $C_D \gg C_{LC}$ is not satisfied. In this case, the maximum of $\varepsilon''_{AP}$ as given by eq. (3) is not half the value of $\varepsilon_s - \varepsilon_\infty$, neither is the position of the peak at frequency $1/(2\pi\tau)$.

**Figure 1h** gives $\varepsilon''_{AP}$ as predicted from the model equivalent circuit of the LC cell. The plots are obtained for $\varepsilon_s = 2.5 \cdot 10^4$, $\tau = 4$ ms, $\varepsilon_\infty = 3$ and $\rho_{LC} = 10^7$ $\Omega$m for all cells, while the thickness ($d_D$) of the surface layer (and thus capacitance $C_D$) and parasitic resistance $R_x$ depend on the type of cell. All the cells of the same type (G, NA or HT) were fitted with the same $R_x$ and the same thickness of surface layers. The capacitance of surface layers was modelled as $C_D = C_0 df_D$, where $f_D = \varepsilon_D/d_D$ and $\varepsilon_D$ is a relative permittivity of the surface layer, for which we assume that it is frequency independent within the measured frequency range. The plots are given for $f_D$ being 5 nm$^{-1}$ and 4 nm$^{-1}$ in the NA and G cells, respectively, while for HT cells $f_D = 0.6$ nm$^{-1}$. These values are physically sensible as one would expect the surface layer to have approximately the same thickness in cells with no thin layer of surfactant as in this case the thin layer is made of LC molecules (NA and G cells), while in HT cells a surface polymer layer is expected to be thicker. By taking $\varepsilon_D \sim 10$, we find $d_D \sim 2-3$ nm in NA and G cells, while in HT cells $d_D \approx 17$ nm.

The position and magnitude of the peak value of $\varepsilon''_{AP}$ are strongly sensitive to $\varepsilon_s$, $\tau$, $\rho_{LC}$ and $f_D$, while the major effect of $\varepsilon_\infty$ and $R_x$ is on the higher frequency dependence of $\alpha$. A reduction in $R_x$ increases the magnitude of $\alpha$ at high frequencies, while the increase in $\varepsilon_\infty$ shifts the peak (the lowest magnitude of $\alpha$) to lower frequencies. The best fit was obtained for $R_x = 600$ $\Omega$ for NA cells, $R_x = 500$ $\Omega$ for HT cells, i.e. by similar values for the cells with ITO electrodes, as expected, and much lower, $R_x = 50$ $\Omega$ for G cells. When we used the PCG model to explain the DSM results, we found that $R_x$ is crucial in fitting the position of the peak value of $\varepsilon''_{AP}$ because $R_x$ can be comparable to $R_{LC}$ (see SM), thus it essentially affects the time constant of charging/discharging the thin layer surface capacitors.

By considering that inaccuracy in the cell thickness and electrodes' surface area can lead to capacitances of cells to differ by up to 10% from ideal values, we conclude that the proposed equivalent circuit gives not only good qualitative but also semi-quantitative agreement with the experiment. The agreement between model predictions and measurements is obtained only if very high low-frequency relative permittivity of N$_F$ phase is assumed, which is approximately four times higher than the maximum of $\varepsilon''_{AP}$ measured in the NA10 cell. The presented model explains why the apparent relative permittivity increases with increasing cell thickness: by increasing the cell thickness, $C_{LC}$ decreases and we are moving towards the limit $C_{LC} \ll C_D$ in which the apparent relative permittivity values would coincide with the actual material parameters (**Figure S3**). We used the obtained parameters to estimate $C_{LC}$ and $C_D$ for different cells and obtain $C_{LC} \approx 0.6 C_D \approx 0.5$ µF for G10 cell, $C_{LC} \approx 4 C_D \approx 2$ µF for HT10 cell, and $C_{LC} \approx 0.5 C_D \sim 3.8$ µF for NA10 cell, thus the capacitances are indeed comparable. To fulfill the condition $C_{LC} \ll C_D$ the cell thickness should be of the order of millimeters, while in standard experiments cells not thicker than approximately 100 µm are used.

To conclude, we have shown that the low frequency (up to 1 MHz) response obtained by DSM of a ferroelectric nematic material filled in thin planar capacitors can be interpreted in terms of the same complex relative permittivity, independent of cell thickness, type of electrodes and surface treatment of electrodes. The standard procedure, in which the complex impedance is calculated from the current through the cell gives only an apparent complex relative permittivity which is related to material properties in a rather comprehensive way. The reason lies in a huge relative permittivity of the LC in the $N_F$ phase, so the capacitance of the $N_F$ layer becomes comparable to the capacitance of a thin surface layer of surfactant, or, in the case of bare electrodes, a thin layer of LC molecules anchored to the electrodes. DSM results are explained by modelling the LC cell with an equivalent circuit presented in **Figure 1f**. By using the same complex relative permittivity for all the studied cells, it is shown that the apparent relative permittivity will increase with increasing cell thickness, while the apparent relaxation frequency does not necessarily behave in a monotonic way, as observed experimentally and predicted by the model (see **Figure 1d,h**, graphs for NA cells, and **Figure S3**). We conclude that the low-frequency relative permittivity ($\varepsilon_s$) of the ferroelectric nematics is indeed huge, and it is even higher than the apparent measured values. Therefore, the interpretation of DSM for strongly polar LC phases (for example, interpretation of the apparent complex relative permittivity as a function of temperature or bias field) should be revisited also within the scope of the here-proposed model. The model might also explain the unusual temperature dependence of the amplitude mode close to the $N_F$ – N transition, reported in Ref. [23].

Acknowledgments

N. V. and V.M. acknowledge the support of the Slovenian Research and Innovation Agency (ARIS), through the research core funding program No. P1-0055 and No. P2-0028, respectively. E.G., D. P. and J. M. acknowledge the support by the National Science Centre (Poland) under the grant no. 2021/43/B/ST5/00240.

Supplemental Materials for:

# Interpretation of dielectric spectroscopy measurements of ferroelectric nematic liquid crystals

Vojko Matko[1], Ewa Gorecka[2], Damian Pociecha[2], Joanna Matraszek[2], Nataša Vaupotič[3,4*],

[1] University of Maribor, Faculty of Electrical Engineering and Computer Science, Koroška 46, 2000 Maribor
[2] University of Warsaw, Faculty of Chemistry, Zwirki i Wigury 101, 02-089 Warsaw, Poland
[3] University of Maribor, Faculty of Natural Sciences and Mathematics, Koroška 160, 2000 Maribor, Slovenia
[4] Jozef Stefan Institute, Jamova 39, 1000 Ljubljana, Slovenia

*corresponding author: natasa.vaupotic@um.si

**Contents:**

# 1 Experimental techniques

In this section we discuss the Bode plot method used to perform dielectric spectroscopy measurements (DSM). The derivations given below are standard calculations known from the elementary course of Physics of measurements in electro engineering. Nevertheless, we give them step by step, because in standard measurements with impedance analyzers one usually does not think about the background "raw" measurements from which the complex relative permittivity is calculated based on some assumed equivalent circuit of the investigated liquid crystal (LC) cell.

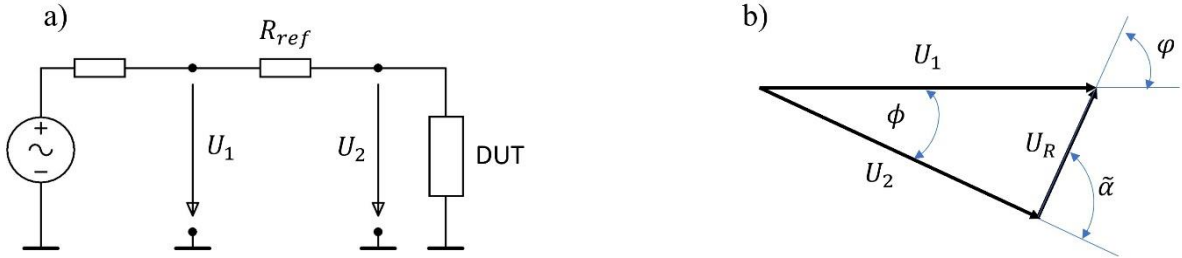

**Figure S1**. a) Electric circuit for the Bode plot method. $U_1$ is given by the Siglent SDG6022X 200MHz Function/Arbitrary Waveform Generator (16 bit), $U_2$ is measured by the Siglent SDS6054A 4CH 500MHz 5 GSa/s Oscilloscope (16 bit), $R_{ref}$ is a resistance of the reference resistor, DUT is a "device under test", in our case a series connection of a LC cell and an ohmic resistor with resistance $R_c$ used for the compensation of the high frequency response of the probes. b) Phasor diagram for the analysis of the Bode plot method: $\phi$ is a phase angle between the applied (generator) voltage ($U_1$) and the voltage ($U_2$) across DUT, $\varphi$ is the phase angle between the current through DUT and the applied voltage, $\tilde{\alpha}$ is a phase angle between the voltage across DUT and current through DUT, which is in phase with the voltage $U_R$ across $R_{ref}$.

The schematic circuit of the **Bode plot method** is shown in **Figure S1a**. We measure the voltage ($U_2$) across the device under test (DUT) and the phase angle ($\phi$) between $U_2$ and voltage ($U_1$) at the generator output as a function of frequency ($\nu$). Usually, a gain $G = 20 \, \text{dB} \log(U_2/U_1)$ is measured, thus $U_2 = U_1 \, 10^{G/20}$.

The current through the cell is obtained by calculating the current through the reference resistor as (see the phasor diagram in **Figure S1b**)

$$I = \frac{1}{R_{ref}} \sqrt{U_2^2 + U_1^2 - 2 U_2 U_1 \cos \phi} \; . \tag{S1}$$

The phase angle ($\varphi$) between the current through the reference resistor ($R_{ref}$) and the applied voltage is

$$\tan \varphi = \frac{-U_2 \sin \phi}{U_1 - U_2 \cos \phi} \tag{S2}$$

and the phase angle between the voltage across DUT and current through DUT is

$$\tilde{\alpha} = \phi - \varphi \; . \tag{S3}$$

The impedance of DUT is $\tilde{Z} = U_2/I$:

$$\tilde{Z} = \frac{R_{ref} \, 10^{\frac{G}{20}}}{\sqrt{10^{\frac{G}{10}} + 1 - 2 \cdot 10^{\frac{G}{20}} \cos \phi}} \; . \tag{S4}$$

We point out that the impedance is independent of the used $R_{ref}$, because $G$ and $\phi$ also depend on $R_{ref}$.

In our measurements, DUT was a series connection of a LC cell and resistor with resistance $R_c$, which was chosen such that a high frequency response of the probes was compensated up to 10 MHz. The measurements on LC cells were reliable up to few MHz, for higher frequencies additional effects appear, related probably to parasitic conductance.

If we interpret the measured $\tilde{Z}$ and $\tilde{\alpha}$ as being a result of a resistor and capacitor in series, the resistance ($R_S$) of the LC cell and capacitance ($C_S$) are obtained as

$$R_S = \tilde{Z}\cos\tilde{\alpha} - R_c \quad , \quad C_S = \left(\tilde{Z}\,\omega\sin\tilde{\alpha}\right)^{-1} , \tag{S5}$$

where $\omega = 2\pi\nu$. The impedance ($Z$) of the LC cell is

$$Z = \sqrt{R_S^2 + \left(\tilde{Z}\sin\tilde{\alpha}\right)^2} \tag{S6}$$

and the angle between the real and imaginary part of $Z$ is

$$\tan\alpha = \frac{\tilde{Z}\sin\tilde{\alpha}}{R_S} . \tag{S7}$$

We used $U_1 = 100$ mVpp in all measurements. The resistance $R_{ref}$ was 9,7 kΩ. The values of $R_c$ were fitted for each channel and each measurement and were from few 100 Ω to few kΩ.

The apparent value of the imaginary part of complex permittivity ($\varepsilon''_{AP}$) is calculated as (eq. (3) in the main text)

$$\varepsilon''_{AP} = \frac{I_0}{\omega C_0 U_{cell}} , \tag{S8}$$

where $I_0$ is the current through the cell in phase with the voltage ($U_{cell}$) across the cell. We note that $I_0 = I\cos\alpha$ and use eq. (S11) to obtain

$$\varepsilon''_{AP} = \frac{\cos\alpha}{\omega C_0 Z} . \tag{S9}$$

## 2 Additional data

In this Section we present the Bode plot measurements of the studied material (**Material 1**) and some additional model results for this material. We also present measurements of another typical $N_F$ material (**Material 2**) exhibiting a direct transition from the isotropic to the ferroelectric nematic phase at 85°C; measurements were performed at 80°C. Results are analyzed within the scope of the PCG model and our model, which we call the "high $\varepsilon$" model.

**Figure S2** gives the raw measured data, gain $G$ and phase angle ($\phi$) between the voltage across DUT (LC cell and $R_c$ in series) and voltage across the generator output for material 1 in the $N_F$ phase. From this data we calculated the impedance ($Z$) of the cell and the angle ($\alpha$) between the current through the cell and voltage across the cell as well as $C_S$ and $R_S$. These results are given in **Figure 1** of the main text and are to be compared with the high $\varepsilon$ model results given in **Figure S3**. Model parameters are collected in
**Table S1**. In **Figure S3** we also give the apparent value of the imaginary part of complex permittivity ($\varepsilon''_{AP}(\nu)$) as predicted by the high $\varepsilon$ model for different cell thicknesses, going beyond the cell

thicknesses used in the experiment. We see that by increasing the cell thickness the magnitude and position of the peak value of $\varepsilon''_{AP}(\nu)$ move towards the frequency $1/(2\pi\tau)$ and magnitude $\varepsilon_s/2$, as expected. **Figure S4** gives the results predicted by the PCG model at the parameter values collected in **Table S2**. We fitted the resistivity of the LC ($\rho_{LC}$) and the thickness of the surface capacitor, which is given by parameter $f_D$ (see the main text). The value of $\varepsilon_s$ was set to 5.0, but it could be also of the other of 10 or 100 and it would not affect the fitting of the PCG model. Because in the PCG model $R_{LC}$ is low, the current will flow only through $R_{LC}$, if frequencies are low enough. This means that the capacitor $C_{LC}$ does not affect the DSM results et all. The values of $\tau$ and $\varepsilon_\infty$ are thus irrelevant, so we kept them the same as obtained from the high $\varepsilon$ model. The parasitic resistance $R_x$ was used also in the PCG model because it turned out to be essential in fitting the positions of the peaks in $\varepsilon''_{AP}$ with the same $\rho_{LC}$ in all cells. We used the same values of $R_x$ as fitted in the high $\varepsilon$ model. Resistances $R_{LC}$ of the cells, calculated from the fitted resistivity as $R_{LC} = \rho_{LC}d/S$, where $d$ and $S$ are the cell thickness and surface area, respectively, are given in **Table S3**. They are rather high, up to 10 kΩ in the G10 cell.

At low frequencies, the PCG model predicts the same capacitances of the cells with the same type and area of the surface electrodes, while the measurements show that these capacitances differ (see Table 1 in the main paper). The high $\varepsilon$ model properly accounts for these differences. The models differ also in the predicted dependency of $\varepsilon''_{AP}(\nu)$ in thick cells, which in the case of the high $\varepsilon$ model settles at $\varepsilon''(\nu)$ of the LC material, while in the PCG model the peak moves towards low frequencies and eventually disappears. Also, in the PCG model, the parasitic resistance $R_x$ is required to fit the position and magnitude of the peaks in $\varepsilon''_{AP}$ with the same resistivity of LC in all cells, while in the high $\varepsilon$ model it is required to model the high frequency response of the cells. In the PCG model the high frequency response is due to the capacitance of the LC layer with some low value of the dielectric constant.

Measurements with material 2 (**Figure S5**) were performed only in three cells (based on the availability of the cells): G10, G5 and NA3. **Figure S6** gives the high $\varepsilon$ model predictions. The model parameters are collected in **Table S1**. For this material, the optimum fit is obtained for the following set of parameter values: $\varepsilon_\infty = 5$, $\varepsilon_s = 3.0 \cdot 10^5$, $\tau = 0.45$ ms and $\rho_{LC} = 10^7$ Ωm. The values of $f_D$ are the same for both the NA and G cells: $f_D = 5$ nm$^{-1}$, the value of $R_x = 50$ Ω for the gold cells is the same as for Material 1, but the model predictions are very weakly dependent on lowering this value below 50 Ω. For the NA3 cell, however, the position of the peak $\varepsilon''_{AP}$ strongly depends on $R_x$ and the best fit is found for $R_x = 200$ Ω, i.e. a few times lower than for material 1 but still of the same order of magnitude. The model predicts a monotonic increase of the maximum value of the apparent imaginary part of complex permittivity with cell thickness, both for the G and NA cells. On the other hand, apparent relaxation frequency decreases towards the LC Debye relaxation frequency with increasing cell thickness in G cells, while in NA cells it increases towards this value.

**Figure S7** gives the PCG model predictions for material 2. Parameter values are collected in **Table S2**. Again, we fitted only the values of $\rho_{LC}$ and $f_D$. The value of $\varepsilon_s$ was set to 5.0, while the rest of parameters were left the same as in the high $\varepsilon$ model. Again, the differences in the low frequency values of capacitance (and impedance) of the cells can be accounted for only by the high $\varepsilon$ model.

**Table S1**. Model parameters for the high $\varepsilon$ model. Their definition is given in the main text.

| material | $\Delta\varepsilon$ [× 10$^4$] | $\tau$ [ms] | $\varepsilon_\infty$ | $\rho_{LC}$ [MΩ m] | $f_s^{(G)}$ [nm$^{-1}$] | $f_s^{(NA)}$ [nm$^{-1}$] | $f_s^{(HT)}$ [nm$^{-1}$] | $R_x^{(G)}$ [Ω] | $R_x^{(NA)}$ [Ω] | $R_x^{(HT)}$ [Ω] |
|---|---|---|---|---|---|---|---|---|---|---|
| 1 | 2.5 | 4.0 | 3 | 10 | 4.0 | 5.0 | 0.6 | 50 | 600 | 500 |
| 2 | 3.0 | 0.45 | 5 | 10 | 5.0 | 5.0 | / | 50 | 200 | / |

**Table S2**. Model parameters for the PCG model. Their definition is given in the main text.

| material | $\Delta\varepsilon$ | $\tau$ [ms] | $\varepsilon_\infty$ | $\rho_{LC}$ [kΩ m] | $f_s^{(G)}$ [nm$^{-1}$] | $f_s^{(NA)}$ [nm$^{-1}$] | $f_s^{(HT)}$ [nm$^{-1}$] | $R_x^{(G)}$ [Ω] | $R_x^{(NA)}$ [Ω] | $R_x^{(HT)}$ [Ω] |
|---|---|---|---|---|---|---|---|---|---|---|
| 1 | 5.0 | 4.0 | 3 | 22 | 2.2 | 2.6 | 0.4 | 50 | 600 | 500 |
| 2 | 5.0 | 0.45 | 5 | 1.5 | 2.8 | 4.1 | / | 50 | 200 | / |

**Table S3.** Resistances ($R_{LC}$) of cells (material 1) calculated from the PCG model.

| cell | G5 | G10 | NA3 | NA5 | NA10 | HT10 | HT20 |
|---|---|---|---|---|---|---|---|
| $R_{LC}$ [kΩ] | 4.4 | 8.8 | 0.40 | 2.0 | 1.3 | 2.5 | 4.7 |

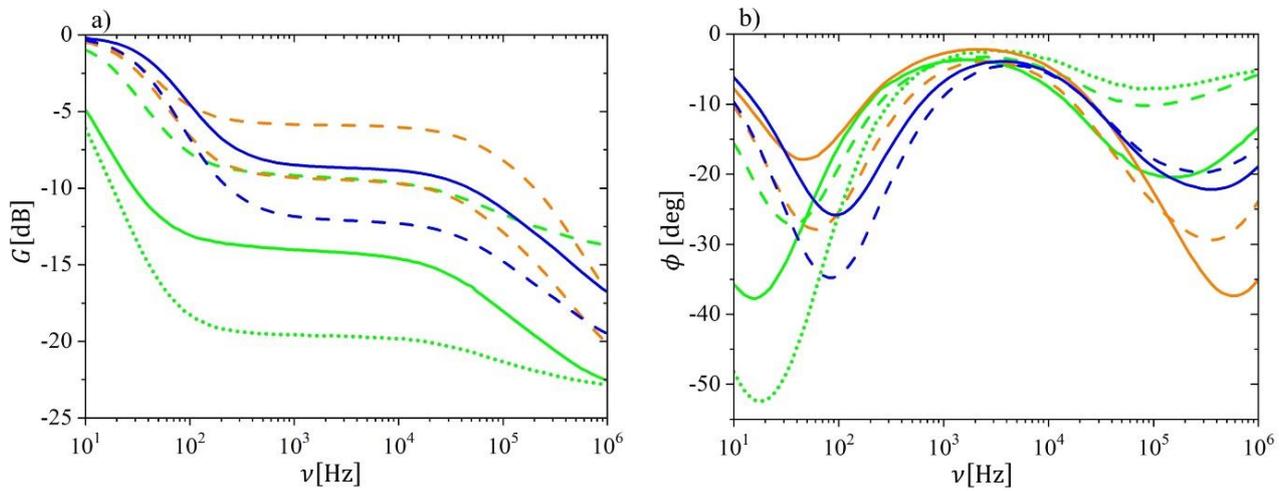

**Figure S2**. **Material 1** in the N$_F$ phase at 49.0°C. **Bode plot measurements.** a) Gain ($G$) and b) phase angle ($\phi$) between the voltage across DUT (LC cell and $R_c$ in series) and voltage on the generator output. Green: NA10 (solid), NA5 (dashed), NA3 (dotted). Yellow: G10 (solid), G5 (dashed). Blue: HT20 (solid), HT10 (dashed).

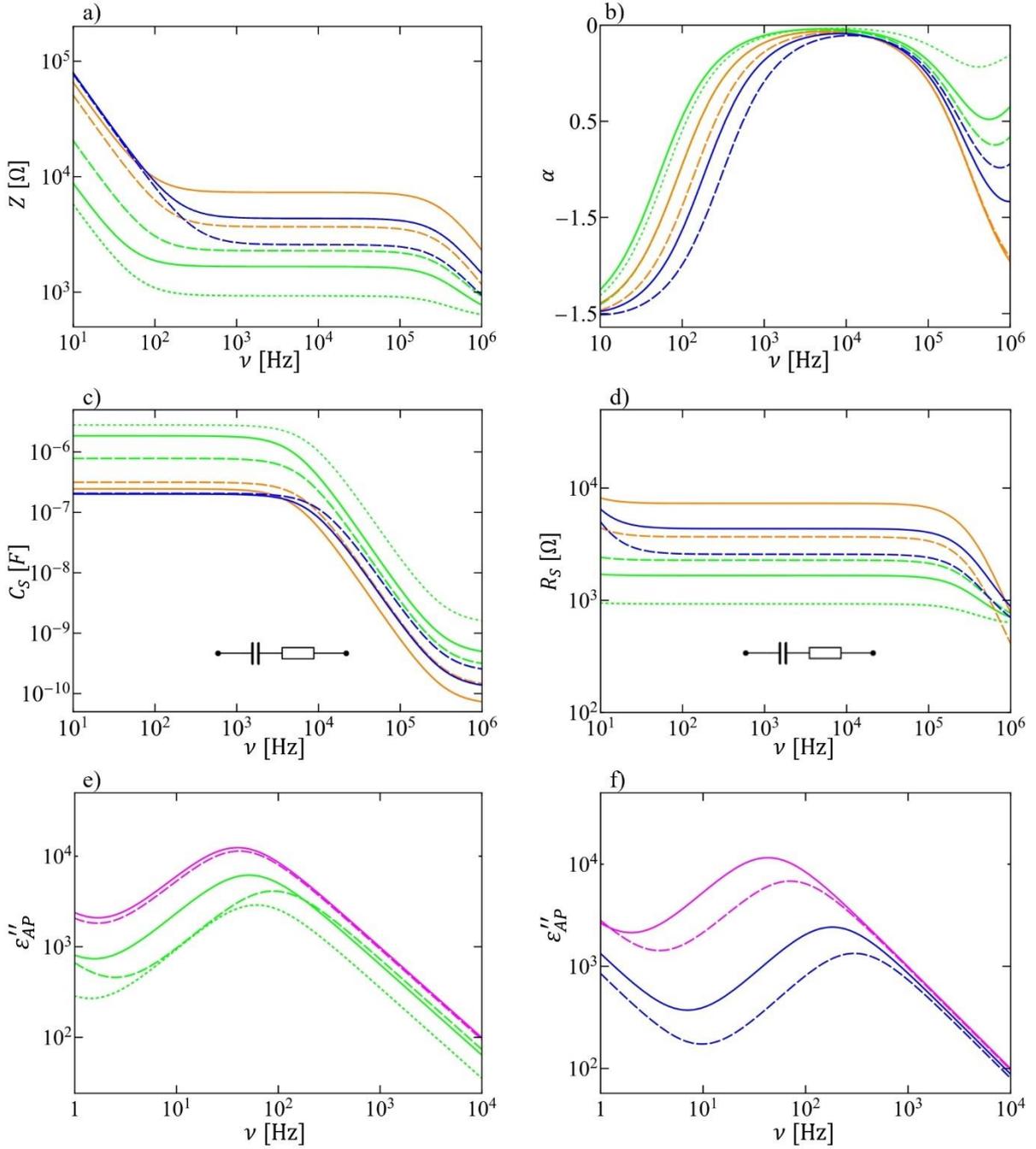

**Figure S3**. **Material 1. Model results (high $\varepsilon$).** a) Impedance ($Z$) and b) phase angle ($\alpha$) between the current through the cell and voltage across the cell as a function of frequency ($\nu$), calculated for the equivalent circuit of the LC cell (Figure 1f) with parameters given in Table S1. Apparent c) capacitance ($C_S$) and d) resistance ($R_S$) of the cell obtained by simulating $Z$ and $\alpha$ of the model electric circuit as if resulting from a series connection of a resistor and capacitor (to be compared with experimental results in Figure 1). Apparent imaginary part of complex permittivity ($\varepsilon''_{AP}$) at different cell thicknesses for the e) NA and f) HT cells. Here the green and blue curves present the model results for the studied cells, and magenta curves for a 100 µm (dashed curve) and 1 mm (solid curve) thick cells. Green: NA10 (solid), NA5 (dashed), NA3 (dotted). Orange: G10 (solid), G5 (dashed). Blue: HT20 (solid), HT10 (dashed).

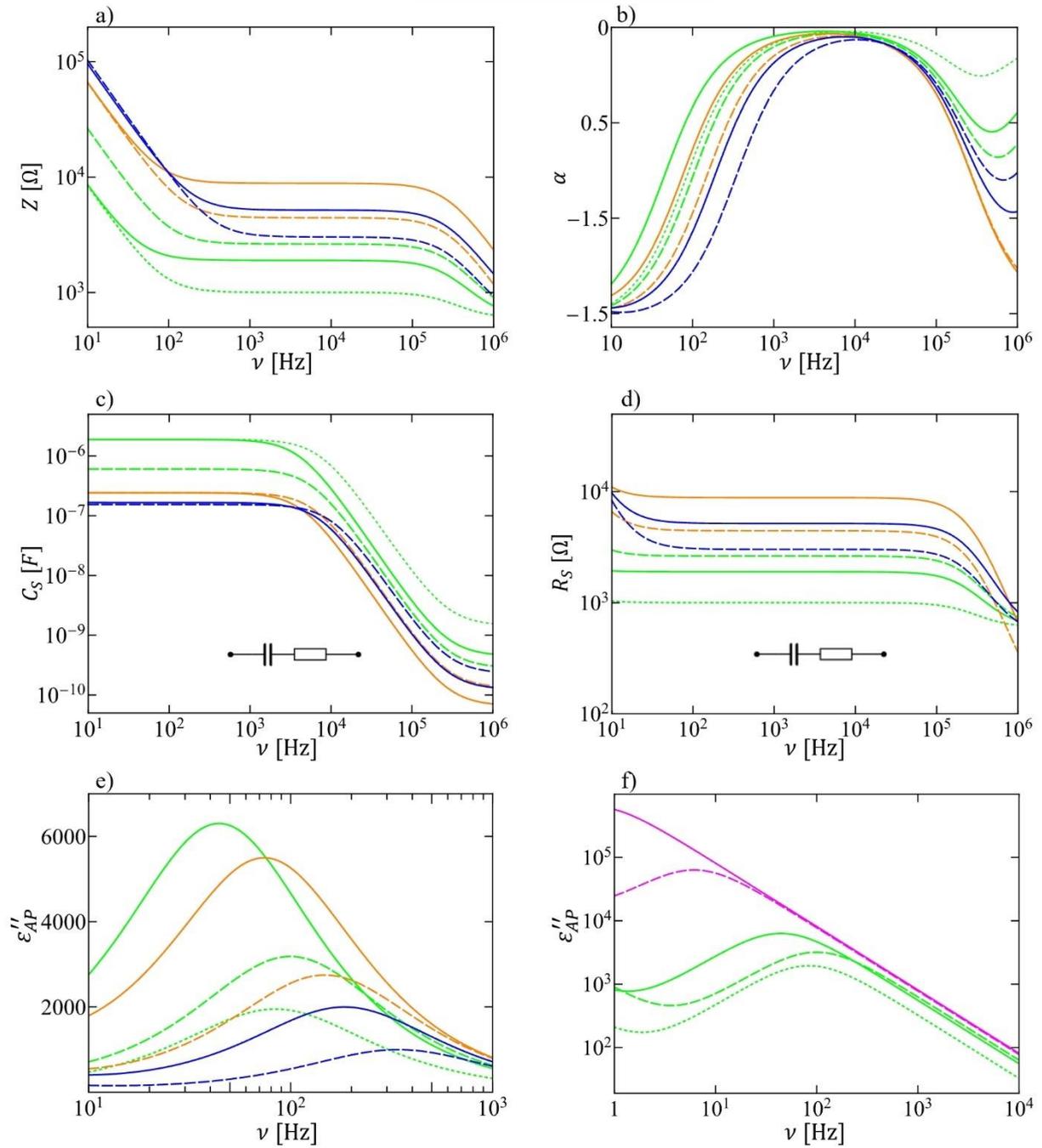

**Figure S4. Material 1. Model results (PCG).** a) Impedance ($Z$) and b) phase angle ($\alpha$) between the current through the cell and voltage across the cell as a function of frequency ($\nu$), calculated for the equivalent circuit of the LC cell (Figure 1f) with parameters given in Table S2. Apparent c) capacitance ($C_S$) and d) resistance ($R_S$) of the cell obtained by simulating $Z$ and $\alpha$ of the model electric circuit as if resulting from a series connection of a resistor and capacitor (to be compared with experimental results in Figure 1). Apparent imaginary part of complex permittivity ($\varepsilon''_{AP}$) for e) all the studied cells and f) at different cell thicknesses of the NA cells. Here the green curves present model results for the studied cells and magenta curves for a 100 μm (dashed curve) and 1 mm (solid curve) thick cell. Green: NA10 (solid), NA5 (dashed), NA3 (dotted). Orange: G10 (solid), G5 (dashed). Blue: HT20 (solid), HT10 (dashed).

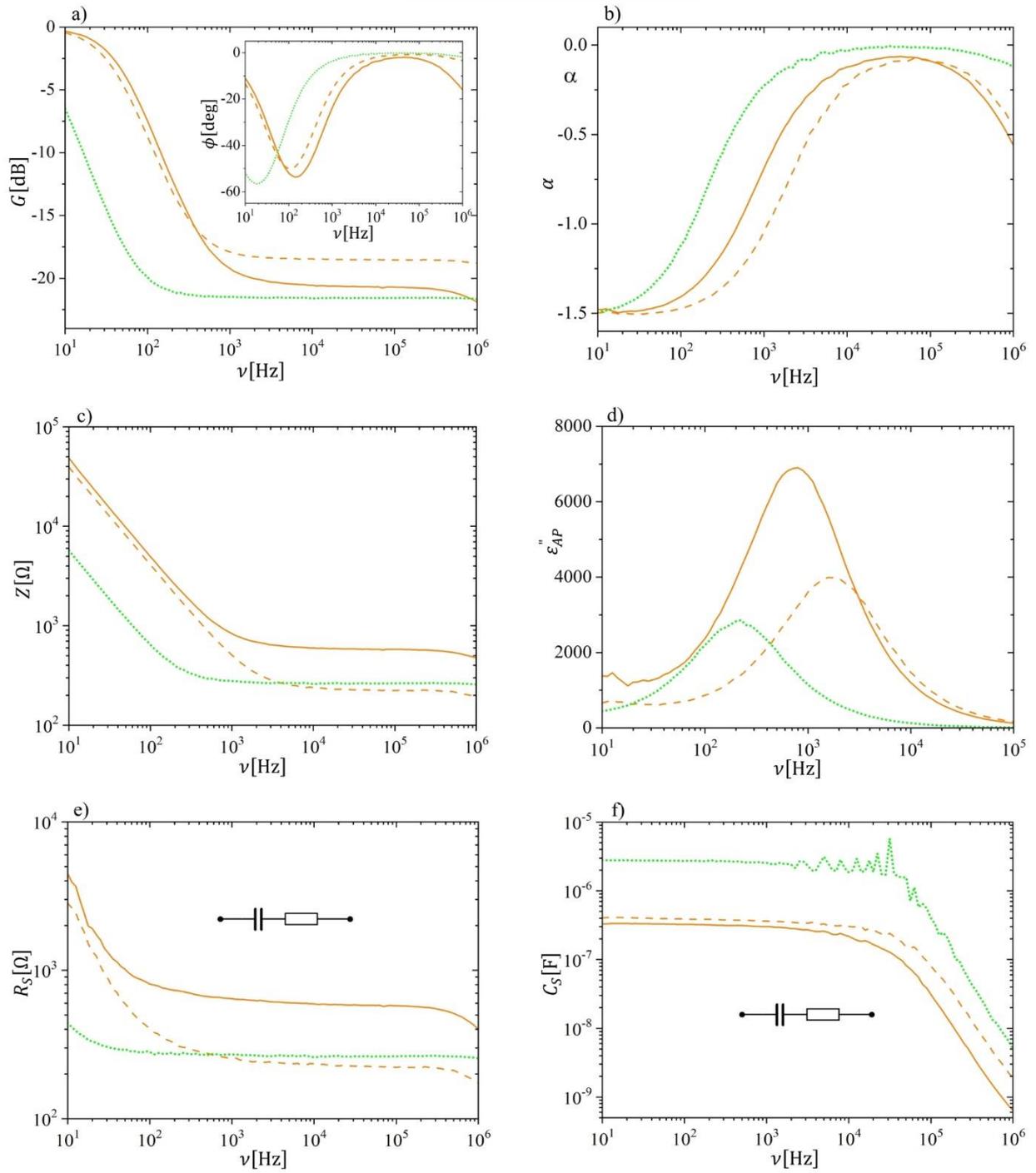

**Figure S5. Material 2** in the N$_F$ phase at 80.0°C. **Bode plot measurements**. a) Gain ($G$) and (inset) phase angle ($\phi$) between the voltage across DUT (LC cell and $R_c$ in series) and voltage across DUT). b) Phase angle ($\alpha$) between the current through the cell and voltage across the cell and c) impedance ($Z$) calculated from $G$ and $\phi$. d) Apparent imaginary part of complex permittivity ($\varepsilon''_{AP}$) calculated by eq.(S14). Apparent e) resistance ($R_S$) and f) capacitance ($C_S$) of the cell as a function of frequency ($\nu$) by simulating the measured $Z$ and $\alpha$ as if resulting from a series connection of an ideal resistor and capacitor. Green: NA3 (dotted). Orange: G10 (solid), G5 (dashed).

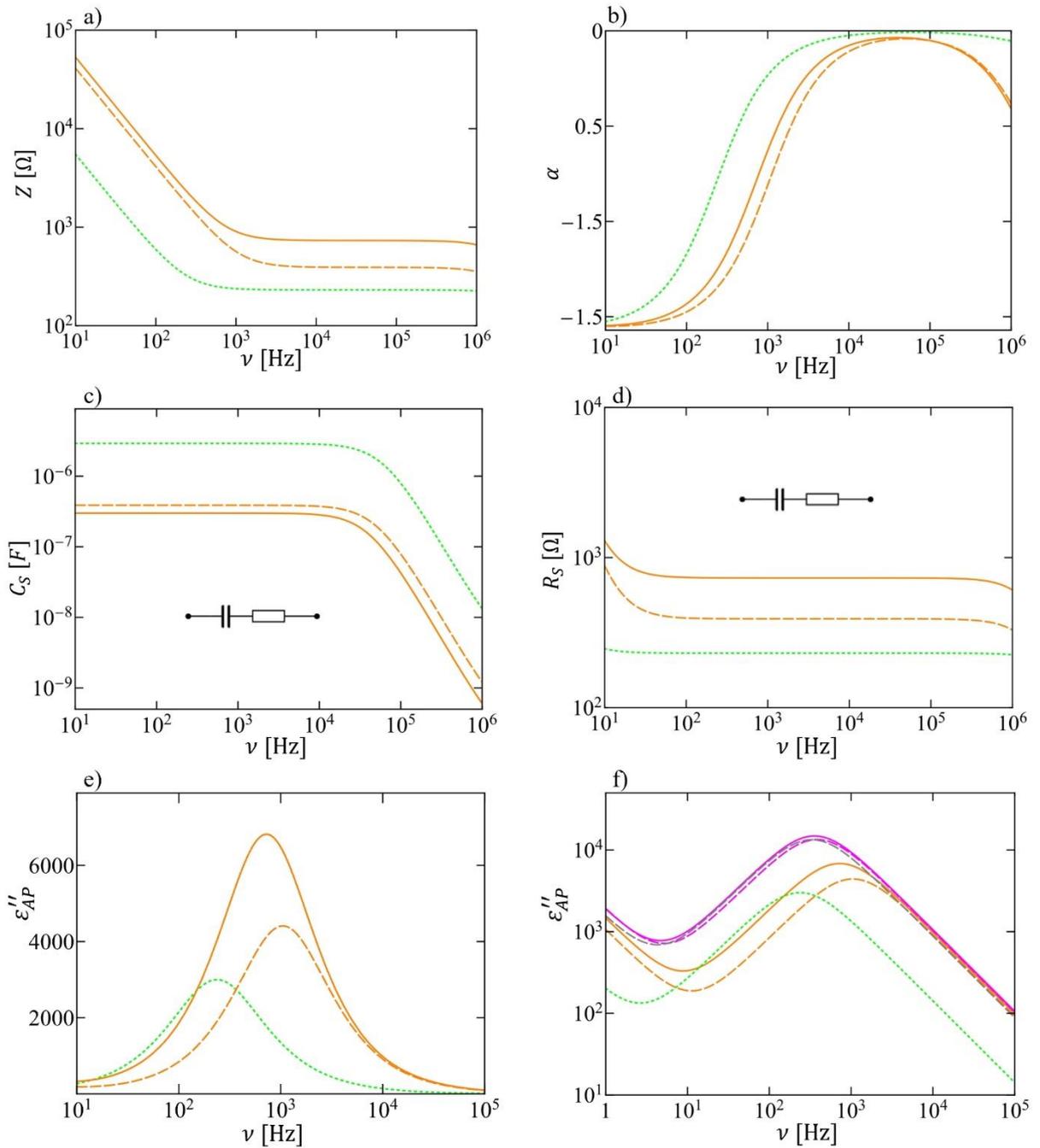

**Figure S6. Material 2. Model results (high $\varepsilon$).** a) Impedance ($Z$) and b) phase angle ($\alpha$) between the current through the cell and voltage across the cell as a function of frequency ($\nu$), calculated for the equivalent circuit of the LC cell (Figure 1f) with parameters given in Table S1. Apparent c) capacitance ($C_S$) and d) resistance ($R_S$) of the cell obtained by simulating $Z$ and $\alpha$ of the model electric circuit as if resulting from a series connection of a resistor and capacitor. Apparent imaginary part of complex permittivity ($\varepsilon''_{AP}$) for e) the measured cells and f) also for thicker cells: 100 μm thick G cell (magenta, dashed), 1 mm thick G cell (magenta, solid) and 100 μm thick NA cell (gray, dashed). Green: NA3 (dotted). Orange: G10 (solid), G5 (dashed).

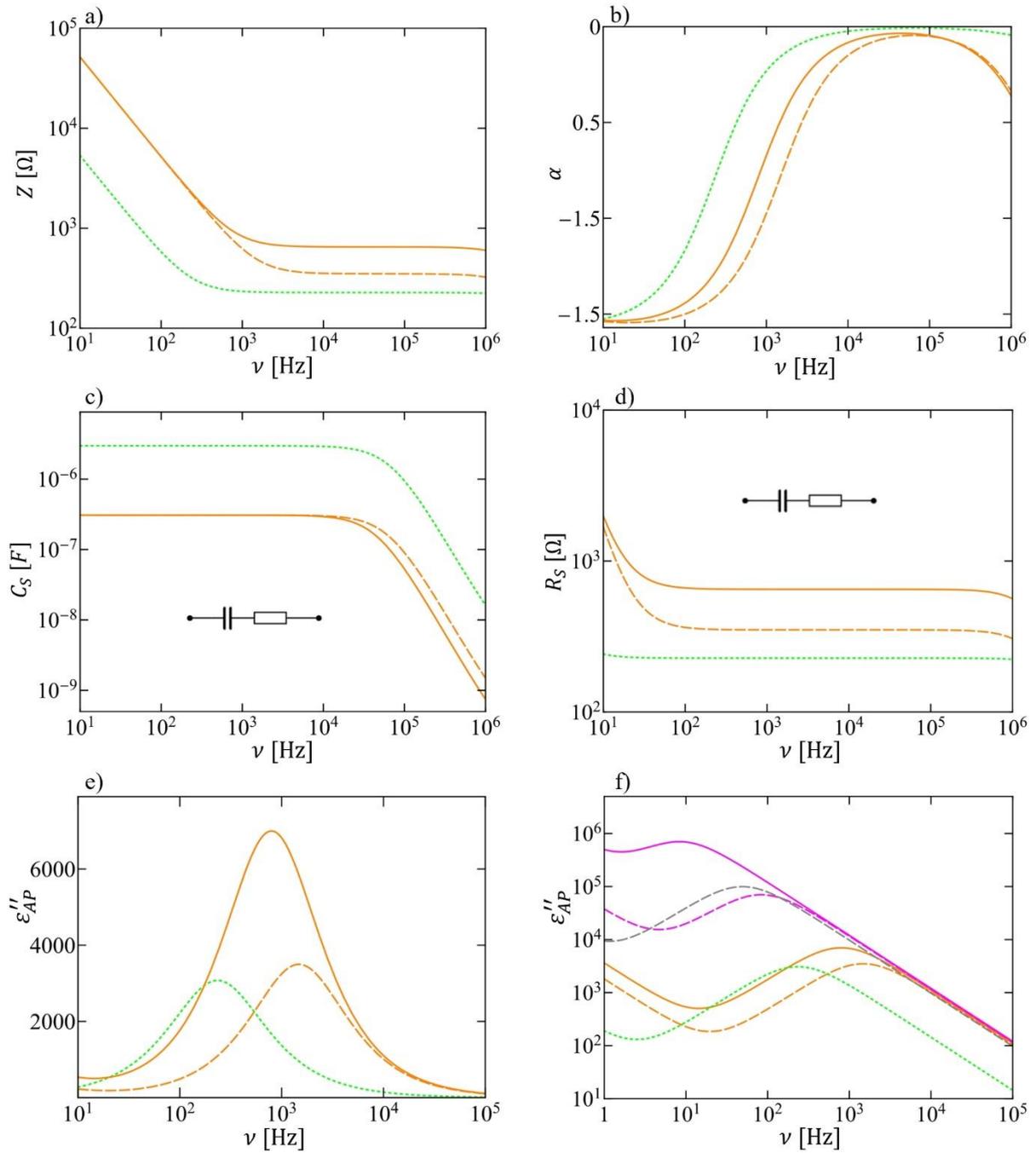

**Figure S7. Material 2. Model results (PCG).** a) Impedance ($Z$) and b) phase angle ($\alpha$) between the current through the cell and voltage across the cell as a function of frequency ($\nu$), calculated for the equivalent circuit of the LC cell (Figure 1f) with parameters given in Table S2. Apparent c) capacitance ($C_S$) and d) resistance ($R_S$) of the cell obtained by simulating $Z$ and $\alpha$ of the model electric circuit as if resulting from a series connection of a resistor and capacitor. Apparent imaginary part of complex permittivity ($\varepsilon''_{AP}$) for e) the measured cells and f) also for thicker cells: 100 µm thick G cell (magenta, dashed), 1 mm thick G cell (magenta, solid) and 100 µm thick NA cell (gray, dashed). Green: NA3 (dotted). Orange: G10 (solid), G5 (dashed).

20